\documentclass[twocolumn,twocolappendix]{aastex631}
\usepackage{xcolor,soul,color}
\usepackage[inline,shortlabels]{enumitem}
\usepackage{amsmath}
\usepackage{bm}
\usepackage{hyperref}
\usepackage{threeparttable}
\newcommand{\MZ}[1]{\textbf{\textcolor{blue}{[M:#1]}}}
\newcommand{\YDX}[1]{\textbf{\textcolor{magenta}{[YD:#1]}}}
\newcommand{\XLC}[1]{\textbf{\textcolor{red}{[XL:#1]}}}

\defcitealias{2025arXiv251118494D}{Paper~I}

\begin{document}

\title{Synthesis imaging with a lunar orbit array: II. Impacts of instrument-induced phase errors}

\correspondingauthor{Meng Zhou}
\email{zhoumeng@bao.ac.cn}

\author[0000-0002-2744-0618]{Meng Zhou}
\affiliation{State Key Laboratory of Radio Astronomy and Technology, National Astronomical Observatories, CAS, 20A Datun Road, Chaoyang District, Beijing 100101, People’s Republic of China}

\author[0000-0001-8075-0909]{Furen Deng}
\affiliation{State Key Laboratory of Radio Astronomy and Technology, National Astronomical Observatories, CAS, 20A Datun Road, Chaoyang District, Beijing 100101, People’s Republic of China}
\affiliation{School of Astronomy and Space Science, University of Chinese Academy of Sciences, Beijing 100049, People’s Republic of China}

\author[0000-0003-3224-4125]{Yidong Xu}
\affiliation{State Key Laboratory of Radio Astronomy and Technology, National Astronomical Observatories, CAS, 20A Datun Road, Chaoyang District, Beijing 100101, People’s Republic of China}

\author{Li Zhou}
\affiliation{National Space Science Center, Chinese Academy of Sciences, Beijing 100190, People’s Republic of China}

\author[0000-0001-6475-8863]{Xuelei Chen}
\affiliation{State Key Laboratory of Radio Astronomy and Technology, National Astronomical Observatories, CAS, 20A Datun Road, Chaoyang District, Beijing 100101, People’s Republic of China}
\affiliation{School of Astronomy and Space Science, University of Chinese Academy of Sciences, Beijing 100049, People’s Republic of China}



\begin{abstract}
A lunar orbit interferometer array suffers from a number of systematics. Beyond systematics induced by the imaging algorithm itself and thermal noise considered in \citet{2025arXiv251118494D} (hereafter Paper I), phase errors due to instrumental inconsistency between receivers, geometric error in baseline determination, and clock synchronization error between satellites will also affect synthesis imaging with the space array. 
In this paper, we model different sources of phase errors and quantify their impacts on all-sky and patchy-sky map-making, respectively, for the ultra-long wavelength sky ($f\lesssim30$ MHz), using the Discovering the Sky at the Longest wavelength (DSL) mission (also known as the Hongmeng mission) as an example. We find that in the scheme of all-sky imaging, the angular power spectrum 
can be suppressed uniformly for various sources of phase errors.
To ensure a reconstruction of large-scale structures with $\gtrsim 95\%$ of the angular power spectrum, the phase error should be controlled below $\sim 12^\circ$ on the random instrumental component, or below $\sim 12^\circ$ for constant deviation, or below $1.1$\,ns on the temporal component. With multiple baseline measurements, the baseline determination errors below $1$\,m can also meet the requirement. 
In the scheme of patchy-sky imaging, the S/N of point source detections does not change significantly, except with instrumental phase errors or at high frequencies. The impact of geometric phase error is relatively stronger in the patchy-sky imaging with higher resolution because longer baselines are used and fewer times of baseline measurements can be averaged over within an integration time.
When scaled with wavelength, these results set the basic reference for instrumental requirements for future space interferometers.
\end{abstract}




\section{Introduction} \label{sec:intro}
Astronomical observations at various frequency bands or with multiple messengers unveil the universe from different perspectives. 
Radio observations below $\sim 120$ MHz are very important for the detection of the redshifted 21 cm signals from the epoch of reionization, cosmic dawn, and dark ages, which are the last remaining periods without substantive observations in the history of our
universe. Besides, the structure of the Milky Way and other radio sources at 
frequencies $\lesssim 30$ MHz have not been studied before. However, ground-based telescopes are almost impossible to detect cosmic signals below 30 MHz due to the strong absorption/ reflection of the ionosphere and severe radio frequency interferences (RFI).

Observations on lunar orbits and the far-side surface can avoid ionosphere effects and RFI. Therefore, many low-frequency astronomical mission concepts based on lunar orbits, including DARE \citep{2017ApJ...844...33B}, DAPPER \citep{2019AAS...23421202B}, PRATUSH \citep{2023ExA....56..741S}, SEAMS \citep{2021AIPC.2335c0005B}, CosmoCube \citep{Artuc2024,Zhu2025},
DEX \citep{2024SPIE13092E..2MB}, and DSL \citep{2021RSPTA.37990566C,chen2023}, or on the far-side surface of the Moon, including FARSIDE \citep{2021RSPTA.37990564B}, Lunar Crater Radio Telescope (LCRT; \citealt{9438165}), LuSee \citep{2023AGUFM.P31B..02B}, ALO \citep{2024AAS...24326401K}, FarView
\citep{2024AdSpR..74..528P} and the Large-scale array for radio astronomy on the farside (LARAF; \citealt{2024arXiv240316409C}) have been proposed and aroused strong interest in recent years. Lunar surface experiments leverage existing imaging methods, data processing tools, and other experiences developed in ground-based radio astronomy. They must also overcome many technical challenges on the engineering level, including array constructions, data transmission, energy supply, and dust proofing. Lunar orbit experiments are more technologically mature but lack tailored imaging algorithms and data processing tools.

The DSL mission \citep{2021RSPTA.37990566C,chen2023} is one of the proposed lunar orbit missions. It includes one mother satellite and nine daughter satellites. The mother satellite collects data from the daughter satellites, performs cross-correlations, and transmits the data back to Earth. One of the daughter satellites is designed to measure the global 21\,cm spectrum at the frequency band of 30 -- 120 MHz. The other eight perform global spectrum measurements and interferometric all-sky imaging observations below 30 MHz. The satellites are designed to orbit the Moon in a linear formation on the same circular orbit with precession. Such orbital motions can generate a three-dimensional distribution of baselines across the range from 100 m to 100 km. 

Radio interferometric imaging using such a lunar-orbital array can be very different from imaging using ground-based arrays. New imaging algorithms are needed to tackle the all-sky field-of-view, non-coplanar moving baselines, the mirror symmetry with respect to the orbital plane, and position-dependent sky blockage by the Moon. \citet{2018AJ....156...43H} has proposed a map-making method by solving the linear mapping equations relating the sky intensity to the visibilities and verified its applicability. \citet{2022MNRAS.510.3046S} applied this method to the DSL with a realistic lunar array configuration and presented preliminary sensitivity estimates taking into account thermal noise from both the bright ultra-long wavelength sky and the anticipated receivers. \citetalias{2025arXiv251118494D} addressed the aliasing effect due to sub-pixel noise, studied the choice of the regularization parameter, and incorporated a more realistic array configuration by including an anticipated ``breathing'' strategy in which the satellites move periodically closer and farther to each other along the orbit.

For an interferometric array to work, the clock of different array elements needs to be synchronized, and the baseline vectors between the array elements also need to be determined. For the orbital array where the relative positions of the satellites are dynamically changing, these are more complicated to implement than the ground-based arrays, where the relative positions are basically fixed except for the rotation of the Earth. In the DSL, the clock of each daughter satellite is synchronized to the clock of the mother satellite via a microwave link, and the length of the baseline is determined at the same time using the dual one-way ranging (DOWR) principle for distance measurements (see details in Section \ref{sec:DORW}). The direction of the baseline vector is measured with respect to fixed stars using the star sensor camera onboard. On each satellite, only a very limited amount of resources (installation space, mass, power, communication bandwidth) can be allocated for this purpose. Therefore, errors in baseline determinations and time synchronizations will inevitably induce geometric and temporal phase errors on the phase term of visibility functions, which will affect the imaging result.

In this work, we will model the geometric, temporal, and instrumental phase errors based on the configuration of the DSL mission. We will use the pixel-averaging linear brute-force in \citetalias{2025arXiv251118494D} and beam-forming (Deng et al. in preparation) map-making methods to do imaging and investigate the impacts of phase errors on imaging quality and the dependencies with respect to the frequency and spatial resolution.

The paper is organized as follows. In Section~\ref{sec:algorithm}, we introduce the simulation setup, including the visibility modeling, the input sky map, and the configuration of satellite orbits, and briefly review two imaging algorithms. In Section~\ref{sec:error_model}, we present our modeling of various sources of phase errors. In Section~\ref{sec:results}, we present the impact of phase errors on the all-sky angular power spectrum as well as on the signal-to-noise ratio of point-source detections, and also give some analytical interpretations for the results we have obtained. Finally, in Section~\ref{sec:conclusion} we draw our conclusions.

\section{Imaging Algorithm and Simulation Setup}
\label{sec:algorithm}
\subsection{Modeling the visibility}
\label{subsec:vis}
We start with the basics of interferometric imaging. The visibility $V_{ij}$ measured by some interferometer array elements $i$ and $j$ on a lunar orbit can be written as (\citealp{2018AJ....156...43H,2022MNRAS.510.3046S}; \citetalias{2025arXiv251118494D})
\begin{equation}
    V_{ij}(t) = \int s(\hat{\mathbf{n}})S_{ij}(\hat{\mathbf{n}})A_{ij}(\hat{\mathbf{n}}){\exp}\left(-2\pi {\mathrm{i} }\hat{\mathbf{n}}\cdot\mathbf{r}_{ij}/\lambda\right)d\hat{\mathbf{n}}^2\,,
    \label{eqn:vis}
\end{equation}
where $\hat{\mathbf{n}}$ is the sky direction, $s(\hat{\mathbf{n}})$ is the sky signal, $S_{ij}(\hat{\mathbf{n}})=S_{i}(\hat{\mathbf{n}})S_{j}(\hat{\mathbf{n}})$ is the shading function, $A_{ij}(\hat{\mathbf{n}})=\sqrt{A_{i}(\hat{\mathbf{n}})A_{j}(\hat{\mathbf{n}})}$ is the power response of the primary beam, $\mathbf{r}_{ij}$ is the baseline vector. 

We model the Moon as a fully opaque sphere and neglect its radiation, diffraction, and reflection. As noted in \citetalias{2025arXiv251118494D} and will be further studied in our subsequent papers, these will not significantly affect the result. Moreover, lunar thermal radiation will not induce phase errors as it is incoherent. The phase errors due to lunar diffraction/reflection should be negligible because the phase differences from the diffracted/reflected waves should be very close to the original incident waves. The shading function $S_{i}$ masks out the contributions from the sky directions blocked by the Moon, i.e. 
\begin{equation}
    S_{i}(\hat{\mathbf{n}}) = 
    \begin{cases}
    0,  &-\hat{\mathbf{n}}\cdot \mathbf{r}_{i}>\sqrt{|\mathbf{r}_{i}|^2-R_{\rm moon}^2} \\
    1, &{\rm else}
    \end{cases},
\end{equation}
where $R_{\rm moon}=1737.1\,{\rm{km}}$ is the radius of the Moon.

The DSL mission proposes to use three orthogonal pairs of short dipole antennas \citep{2021RSPTA.37990566C}. So the primary beam of the antenna $i$ can be approximated as \citep{2022MNRAS.510.3046S}:
\begin{equation}
    A_i(\hat{\mathbf{n}})=1-(\hat{\mathbf{n}}\cdot\hat{\mathbf{n}}_i^a)^2\,,
\end{equation}
where $\hat{\mathbf{n}}_i^a$ is the antenna direction. In this work, we follow the assumption in \citet{2018AJ....156...43H}, \citet{2022MNRAS.510.3046S}, and \citetalias{2025arXiv251118494D} that the beam null of each antenna always points to the center of the Moon.

The thermal noise $\eta_{ij}$ on the measured visibility $V_{ij}$ can be modeled as white noise. Its real part and imaginary part should follow independent Gaussian distributions with a standard deviation given by
\begin{equation}
    \sigma_{ij}=\frac{T_{ij}^{\rm sys}}{\sqrt{2\Delta\nu t_{\rm int}}}\,,
\end{equation}
where $T_{ij}^{\rm sys} = \sqrt{(T^{\rm rec}+T_i^{\rm sky})(T^{\rm rec}+T_j^{\rm sky})}$. Here we assume the receiver temperature $T^{\rm rec}$ is identical for each antenna. $T_i^{\rm sky}$ is the global sky temperature observed by the antenna $i$. We also neglect the slight differences between $T_i^{\rm sky}$ and $T_j^{\rm sky}$. The effective system temperature can be approximated as $T^{\rm rec}+T_i^{\rm sky}$. The bandwidth is set at $\Delta\nu=8\,{\rm kHz}$. We use the same criteria in paper I to dynamically determine the integration time $t_{\rm int}$\footnote{In practice, the integration time duration on board the satellite is fixed. However, during the post-processing on ground, we can rebin and average visibility data over the dynamic $t_{\rm int}$, which compresses the data without losing information, and the thermal noise level scales down accordingly.}. We keep the change in baselines $|\delta \textbf{r}_{ij}|=\lambda/8$ during $t_{\rm int}$, assuming the relative velocity of satellite $i$ and $j$ is constant between two contiguous time points. The maximum damping to the visibility $V_{ij}$ is ${\rm sinc(\pi/8)}\approx0.974$, which can be neglected. We also set an upper limit $t_{\rm int}<25$ s to avoid significant change of $S(\hat{\mathbf{n}})$ and $A(\hat{\mathbf{n}})$.

\subsection{All-Sky Imaging}
We use the {\it{pixel-averaging}} linear brute-force map-making method in \citetalias{2025arXiv251118494D} to do all-sky imaging.
In \citetalias{2025arXiv251118494D}, we find that the {\it{pixel-averaging}} method, i.e., sampling the beam matrix at a higher resolution and then averaging to the target resolution, can mitigate the aliasing effect due to sub-pixel noise. We also find that including
all baselines shorter than four times the Nyquist limit is sufficient to make maps of good quality. The impact of the regularization parameter is also investigated. Its value can be chosen as a trade-off between the reconstruction error and signal-to-noise ratio. Therefore, intermediate values ($10^{-6}\sim10^{-4}$) are empirically recommended.

We first review the imaging algorithm proposed in \citetalias{2025arXiv251118494D}. For convenience, we rewrite the visibility with thermal noise in matrix form:
\begin{equation}
\label{eqn:vis_matrix}
    \mathbf{V} = \mathbf{B}\mathbf{s}+\bm{\eta}\,,
\end{equation}
where $\mathbf{B}$ is the beam matrix including the shading function, the power response, and the total phase in Equation~(\ref{eqn:vis}).

The reconstructed all-sky map $\mathbf{\hat{s}}$ can be obtained by solving the ill-posed linear Equation~(\ref{eqn:vis_matrix}) via Tikhonov regularization \citep{2017MNRAS.465.2901Z,2024RAA....24b5002Y}, which minimizes the following quadric function
\begin{equation}
    L = (\mathbf{B}\hat{\mathbf{s}}-\mathbf{V})^T\mathbf{N}^{-1}(\mathbf{B}\hat{\mathbf{s}}-\mathbf{V})+(\hat{\mathbf{s}}-\mathbf{s}_p)^T\mathbf{R}(\hat{\mathbf{s}}-\mathbf{s}_p),
\end{equation}
where $\mathbf{N}=\langle\mathbf{\bm{\eta}^\dagger \bm{\eta}}\rangle$ is the noise covariance matrix, $\mathbf{R}$ is the regularization matrix, and $\mathbf{s}_p$ is a prior sky map. The solution is given by
\begin{equation}
    \hat{\mathbf{s}} = (\mathbf{B}^T\mathbf{N}^{-1}\mathbf{B}+\mathbf{R})^{-1}\mathbf{B}^T\mathbf{N}^{-1}(\mathbf{V}-\mathbf{B}\mathbf{s}_p)+\mathbf{s}_p\,.
\end{equation}

We consider a no-prior solution, i.e. $\mathbf{s}_p=0$, and a uniform regularization, i.e. $\mathbf{R}\propto\mathbf{I}$. Then, the so-called ``dirty map'' $\mathbf{s}^{\rm dirty}$ is given by $\mathbf{B}^T\mathbf{N}^{-1}\mathbf{V}$, and the so-called ``clean map'' $\mathbf{s}^{\rm clean}$ is obtained by deconvolving $\mathbf{s}^{\rm dirty}$ with the ``dirty beam'' $\mathbf{B}^T\mathbf{N}^{-1}\mathbf{B}$. 

To get the deconvolution matrix $(\mathbf{B}^T\mathbf{N}^{-1}\mathbf{B}+\mathbf{R})^{-1}$, we first perform eigendecomposition on $\mathbf{B}^T\mathbf{N}^{-1}\mathbf{B}$,

\begin{equation}
    \mathbf{B}^T\mathbf{N}^{-1}\mathbf{B} = \mathbf{Q}\mathbf{W}\mathbf{Q^T}\,,
\end{equation}
where $\mathbf{Q}$ is the column matrix of eigenvectors and $\mathbf{W}$ is the diagonal matrix whose diagonal elements are the corresponding eigenvalues. We use the maximum eigenvalue $W_{\rm max}$ to parameterize $\mathbf{R}$ as follows,
\begin{equation}
    \mathbf{R}= \epsilon W_{\rm max}\mathbf{I}\,,
\end{equation}
where $\epsilon$ is the regularization parameter.
Therefore, the inversion of $\mathbf{B}^T\mathbf{N}^{-1}\mathbf{B}+\mathbf{R}$ can be given by, 
\begin{equation}
     (\mathbf{B}^T\mathbf{N}^{-1}\mathbf{B}+\mathbf{R})^{-1} = \mathbf{Q}(\mathbf{W}+\mathbf{R})^{-1}\mathbf{Q}^{T}\,.
\end{equation}

In principle, $\mathbf{s}^{\rm clean}$ should be the final reconstructed map. But in the case of imperfect deconvolution for a lunar orbit array with limited $uvw$ coverage, we find a strong coupling between the impacts of phase errors and systematics from deconvolution, which makes clean maps less reliable for analysis (see Appendix~\ref{appendix_a} for details).
Therefore, in this work, we mainly focus on the level of dirty maps. The value of $\epsilon$ does not affect the results of this paper as it is not a required parameter for the construction of dirty maps. Without additional clarification, the reconstructed sky maps mentioned in this paper hereafter should be regarded as referring to the dirty maps.

\subsection{Patchy-Sky Imaging}
The impacts of instrument-induced phase errors may depend on the resolution of the images. Due to the limited resolution achievable with the brute-force imaging algorithm, we use the beam-forming algorithm to do patchy-sky imaging in order to achieve the full resolution enabled by the 100 km baseline.
The beam-formed sky temperature at the direction of $\hat{\mathbf{n}}$ can be written as 
\begin{equation}
\label{eqn:beamforming}
    T(\hat{\mathbf{n}}) =\frac{\sum_\alpha w_\alpha V_\alpha \exp(2\pi{\mathrm{i} }\hat{\mathbf{n}}\cdot\mathbf{r}_{\alpha}/\lambda)}{\sum_\alpha w_\alpha S_\alpha(\hat{\mathbf{n}})A_\alpha(\hat{\mathbf{n}}) \Delta \Omega(\hat{\mathbf{n}})}\,,
\end{equation}
where the subscript $\alpha$ represents the sequence of all data points formed by different baselines at different time points. 
Here $w_\alpha$ is the weight factor, which is set to be inversely proportional to baseline counts in the present work, $V_\alpha$ is the visibility function measured by the baseline $\mathbf{r}_{\alpha}$, 
and $\Delta \Omega(\hat{\mathbf{n}})$ is the solid angle. 
If $uvw$ coverage is complete, it can be approximated as $s(\hat{\mathbf{n}})\delta(\hat{\mathbf{n}})$ where $\delta(\hat{\mathbf{n}})$ is the Dirac delta function.

\subsection{Sky Map}
In order to simulate the observed visibility function, we use the same input sky map as described in \citetalias{2025arXiv251118494D}. In brief, we use the Ultra Long wavelength Sky model with Absorption (ULSA; \citealt{2021ApJ...914..128C}) as the diffuse component and generate a mock catalog of point sources based on the source counts of the GaLactic and Extragalactic All-sky MWA survey (GLEAM; \citealt{2019PASA...36....4F}).
ULSA uses a cylindrical emissivity derived from the Haslam map \citep{2015MNRAS.451.4311R} at 408 MHz and extrapolated down to 1 MHz using a power law with respect to frequencies. It also includes free-free absorption by electrons based on the {\tt NE2001} model \citep{2002astro.ph..7156C,2003astro.ph..1598C}.
In this work, we pixelize the sky with the
{\tt HEALPix} pixelization scheme \citep{2005ApJ...622..759G,2019JOSS....4.1298Z}. It can divide a sky map into 12$\times {\rm NSIDE^2}$ quadrilaterals with varying shape but exactly equal area. In the following, we use the parameter NSIDE to define the spatial resolution.

As for the point source catalog, we use GLEAM data at 200, 154, 118, and 88\,MHz \citep{2019PASA...36....4F} to derive a distribution and generate a sample of point sources with random positions. To avoid double counting the point sources that might already exist in the diffuse component, we remove the faint point sources whose fluxes are lower than 
the extrapolated confusion limit of the Haslam map (i.e. 2 Jy at 408 MHz, see paper I for details) with a spectral index of 0.8 \citep{2019PASA...36....4F}.  The remaining catalog contains $\sim$ 2,700 bright point sources and contributes $\sim$ 70\% of the total angular power spectrum of point sources, so the impact of this cutoff on the small-scale power is not significant.


\subsection{Array configurations}
We adopt the designed array configuration for the DSL as in \citetalias{2025arXiv251118494D}.
The orbit of the whole satellite array is modeled as a combination of a uniform circular motion at a height of 300\,km and an inclination angle of 30$^\circ$, and a uniform precession of the orbital plane with a period of 1.3 years. All satellites are assumed to move on the same orbit without deviation.
We only consider the observation time when the Moon blocks the Earth so that we can avoid the RFI from the Earth. 
To obtain a better {\it uvw} coverage and correct for natural velocity deviation due to the inhomogeneous gravitational potential in lunar orbit, a {\it breathing} strategy will be applied, in which the eight daughter satellites move closer and further away from each other periodically covering baselines from 100 m to 100 km. 
The {\it breathing} strategy is quite flexible, provided that the formation flying safety is guaranteed, and can be optimized for different science goals. In this work, we use the same breathing strategy as in paper I, 
and assume that the satellite array will go through two phases during a {\it breathing} period of 14 days.
It is compressed with an overall ratio of $\mathcal{R}$ during the first phase and then recovered to the initial state during the second phase. 
During each of the two phases, each satellite is assumed to have a constant speed and an instantaneous velocity change is applied when changing from one phase to the next.

\begin{table}[]
    \centering
    \caption{The positions of the satellites with respect to the first one at the beginning of each phase.}
    \begin{threeparttable}
    \begin{tabular}{c|c}
    \hline
      Time (day) & $r_n$ \\
      \hline
        0 &  $r^0_n$ \\
        7 &  $r^0_n/\mathcal{R}$ \\
        14 &  $r^0_n$ \\
        \hline
    \end{tabular}
    \begin{tablenotes}
    \footnotesize
    \item \textbf{Note.} Here $r_n^0$ is the initial relative position of the satellite $n$ with respect to the satellite 1.
    \end{tablenotes}
    \end{threeparttable}
    \label{tab:breathing}
\end{table}

We 
denote the initial distance of satellite $n$ to satellite 1 as $r_n^0$ ($n = 1, 2, ..., 8$).
In Table~\ref{tab:breathing}, we list the relative positions of the satellites with respect to the first one at the beginning of each phase for a better illustration.
We adopt the same parameterization of $r_n^0$ as in \citetalias{2025arXiv251118494D}, i.e.
\begin{equation}
\label{eqn:initial_position}
    r_n^0=b_0+a_0q_0^{n-2}\ \ \ \ \  {\rm for}\ \ n\geq2\,.
\end{equation}

We set $r_1^0$ to be 0\,m and $r_8^0$ to be 100\,km. To avoid the risk of collision, we also set the shortest baseline during one {\it breathing} period to be 100\,m and the second shortest one to be 500\,m. Therefore, given the compression ratio $\mathcal{R}$, one can get the parameters $(a_0,\ b_0,\ q_0)$ by plugging the above requirements into Eq.~(\ref{eqn:initial_position}) \citepalias{2025arXiv251118494D}. In this work, we adopt an array configuration with $\mathcal{R}=10$. We have tested that within the formation flying safety constraint, the baseline distribution does not change significantly for different values of $\mathcal{R}$ after a full precession period, as long as the compression ratio is not too small. Therefore, we do not expect a strong dependence of our results on $\mathcal{R}$ within the formation flying safety constraint. 

\section{The Error Model}
\label{sec:error_model}
As mentioned in Section~\ref{sec:intro}, the phase errors of orbital interferometry experiments arise from three kinds of sources, which includes the time synchronization error, the geometric phase error from baseline determination, and the error in determination of the 
instrumental phase of receiver. All these three affects the phase of the visibility functions, their impacts on the interferometer imaging are similar but not exactly the same. In this section, we describe how we model the three kinds of phase errors. Below, we first introduce the principle of the clock synchronization and ranging measurements, and the error in this process, then we proceed to discuss each error in turn.

\subsection{The DOWR principle}
\label{sec:DORW}
Both the clock synchronization and the range measurement between two satellites are accomplished with the dual one-way ranging (DOWR) principle. It requires both satellites to be equipped with reasonably stable and precise clocks, but may have unknown offsets.  
The two satellites A and B transmit signals to each other, and the time interval between signal sending and receiving is used to determine the distance between the two satellites (baseline length) and the offset value of the two clocks, so that the clocks could be synchronized. 

Suppose that A sends out a signal when its local clock reads $t_A = 0$, at this moment, the local clock on B reads $t_B = -t^{\rm off}$, which is the time offset between A and B. When B receives the signal from A, its local clock should read
\begin{equation}
    T_B = t_{tA}+
    D/c+t_{rB}-t^{\rm off}\,,
\end{equation}
where $t_{tA}$ is the transmit delay of A, $D$ is the relative distance between A and B, and $t_{rB}$ is the receive delay of B.

Similarly, B also sends out a signal when its clock reads $t_B = 0$. At this moment, the local clock on A reads $t_A = t^{\rm off}$. When A receives the signal from B, its local clock should read
\begin{equation}
    T_A = t_{tB}+
    D/c+t_{rA}+t^{\rm off}\,,
\end{equation}
where $t_{tB}$ is the transmit delay of B and $t_{rA}$ is the receive delay of A.

In principle, all transmit and receive delays could be determined separately in laboratories. Therefore, $t^{\rm off}$ and $D$ can be solved from $T_A$ and $T_B$ as
\begin{eqnarray}
    D &=& \frac{c}{2}[(T_A+T_B)-(t_{tA}+t_{tB})-(t_{rA}+t_{rB})]\nonumber\\
    t^{\rm off} &=& \frac{1}{2}[(T_A-T_B)-(t_{tB}+t_{rA})+(t_{tA}+t_{rB})]\,.
\end{eqnarray}

The baseline determination errors $\delta D$ and time synchronization errors $\delta t^{\rm off }$ mainly come from measurements of $T_A$ and $T_B$ due to signal pulse detection. Assuming measurement errors $\delta T_A$ and $\delta T_B$ follow an independent and identical distribution, with a zero mean and a variance of $\sigma_t$, the covariance between $\delta D$ and $\delta t^{\rm off}$ is zero, i.e. $\delta D$ and $\delta t^{\rm off}$ are uncorrelated. The magnitudes of $\delta D$ and $\delta t^{\rm off}$ are then
\begin{eqnarray}
    \delta D \approx \frac{c}{2}(\delta T_A+\delta T_B)\approx c\sqrt{2}\sigma_t\nonumber\\
    \delta t^{\rm off} \approx \frac{1}{2}(\delta T_A-\delta T_B)\approx \sqrt{2}\sigma_t
\end{eqnarray}


\subsection{The temporal phase error}
The temporal phase errors $\Delta \phi^{\rm temporal}$ result from inaccuracy in time synchronizations between the mother satellite and the daughter satellites.
The temporal phase error is related to the time synchronization error by
\begin{eqnarray}
    \Delta \phi^{\rm temporal}_{ij} = 2\pi\, \nu \Delta t^{\rm off}_{ij},
\end{eqnarray}
where $\Delta t^{\rm off}_{ij}$ is the error in time offsets between the satellite $i$ and $j$. Each daughter satellite is synchronized with the mother satellite, so that 
$\Delta t^{\rm off}_{ij} = \Delta t^{\rm off}_{i}-\Delta t_{j}^{\rm off}\,$.
We model $\Delta t^{\rm off}_{i}$ as a Gaussian error with a half-width at half-maximum (HWHM) of $\Delta t_0$, where $\Delta t_0$ is the amplitude of the time synchronization error. 

The time offsets are used to update the clocks immediately, so this error induces an error in the phase directly. Due to such an error, the signals from different daughter satellite receivers lose some of their coherence. Within $t_{\rm int}$, the overall impact of temporal phase errors should be evaluated based on the average value of temporal phase shift, i.e.
\begin{equation}
\label{eqn:time_err}
    V^{\rm err}_{ij}=V^{\rm true}_{ij}\overline{\exp(-2\pi{\mathrm i}\nu \Delta t^{\rm off}_{ij})}\,.
\end{equation}
Here we have assumed that $V^{\rm true}$ (and the true voltage data) does not vary during $t_{\rm int}$ and the thermal noise in the voltage data does not correlate with the temporal error.


\subsection{The geometric phase error}

The geometric phase errors $\Delta\phi^{\rm geo}$ arise from the measurement errors in the baseline vector $\mathbf{r}_{\alpha}$.
The distance error results from the timing error when the time delay is measured between the mother satellite and the daughter satellites using the DOWR principle, and the direction measurement error is due to errors in star map comparison using the star sensor camera. In combination, these two sources of errors result in a baseline error of $\Delta \mathbf{r}_{\alpha}$.
In the case of DSL, the direction of a daughter satellite with respect to the mother satellite will be measured with an error of about 2 arcseconds, which is designed in accordance with a distance measurement error of about 1 m at 100 km, forming an error sphere as illustrated in Figure~\ref{fig:baseline determination}.
Therefore, $\Delta\phi^{\rm geo}$ should be incorporated in the beam matrix $\mathbf{B}$ used for imaging after observations, i.e., 
\begin{eqnarray}
    B_{\alpha\beta}^{\rm err} = B_{\alpha\beta}^{\rm true}{\exp}({{\mathrm i}\,\Delta\phi^{\rm geo}_{\alpha\beta}}).
\end{eqnarray}
Here 
$\beta$ denotes the index of the sky direction $\hat{\mathbf{n}}$, $B^{\rm true}_{\alpha\beta}$ is the true beam matrix element, $B^{\rm err}_{\alpha\beta}$ is the beam matrix element with geometric phase errors which will be used for imaging, and $\Delta\phi^{\rm geo}_{\alpha\beta}$ is the geometric phase error given by
\begin{eqnarray}
    \Delta\phi^{\rm geo}_{\alpha\beta} = 2\pi\,\hat{\mathbf{n}}\cdot\Delta \mathbf{r}_{\alpha}/\lambda\,,
\end{eqnarray}
where $\Delta \mathbf{r}_{\alpha}$ is the baseline determination error on $\mathbf{r}_{\alpha}$.
Note that the geometric phase errors have a projection effect. While $\Delta\phi^{\rm geo}_{\alpha\beta}$ is direction-dependent, the overall effect on the imaging is expected to be isotropic after averaging the whole sky.

In the DSL mission, the baseline vectors $\mathbf{r}_{\alpha}$ between the element pair $i$ and $j$ are obtained by evaluating the differential of position vectors measured from the mother satellite to the daughter satellites $i$ and $j$. Therefore, the error on $\mathbf{r}_{\alpha}$ should be,
\begin{equation}
    \Delta\mathbf{r}_{\alpha} = \Delta\mathbf{r}_{i}-\Delta\mathbf{r}_{j}\,.
\end{equation}
We generate random Gaussian errors with a HWHM of $\Delta r_0$ and add them onto the position vector of the satellite $i$ at each time step of measurements. Note that although $\Delta\mathbf{r}_i$ should be independent for all satellites, $\Delta\mathbf{r}_{\alpha}$ should be correlated with any other baselines formed by the satellite $i$ or $j$. We denote this kind of error model by the ``Single'' model. 

\begin{figure}
    \centering
    \includegraphics[width=\columnwidth]{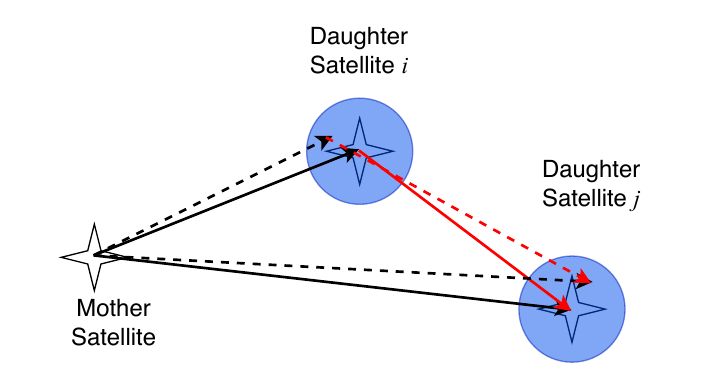}
    \caption{Illustration of the ``Single'' model of baseline measurement errors. The left star represents the mother satellite, while the middle and right stars represent the daughter satellite $i$ and $j$, respectively. The black arrows represent the position vectors from the mother satellite to the daughter satellites, while the red arrows represent the baseline vectors between the daughter satellite $i$ and $j$. The solid arrows represent the vectors without errors, while the dashed ones represent the vectors with errors. The random errors of position vectors are distributed inside the blue spheres.}
    \label{fig:baseline_determination}
\end{figure}

Before being used for imaging, the measurement accuracy of baseline vectors can potentially be further improved by going through some post-processing, making use of the prediction from 
the orbital equations with multiple measurement data and iterative calibrations
\citep{doolittle2005relative,chavez2004relative}. 
Such post-processing can reduce the effective error magnitude but may induce slight correlations on residual errors in limited measurements adjacent in time. Such correlations should not have significant impacts on the reconstructed maps because the effective error magnitude is much lower and the correlations will be damped quickly in time.
We assume that the effective residual error could be similar to the one after averaging over multiple measurements within the integration time duration $t_{\rm int}$. If the baseline vectors are measured every interval of $t_{\rm m}$, then within $t_{\rm int}$, one can average over $N_{\rm eff} = [t_{\rm int}/t_{\rm m}]$\footnote{$[\cdot  ]$ means rounding.} times of measurements. We adopt $t_{\rm m}=0.1$ s as is reasonable for the current technology. Theoretically, the error $\Delta\mathbf{r}_{i}$ should be reduced to $\Delta\mathbf{r}_{i}/\sqrt{N_{\rm eff}}$. If there is no measurement in the integration time duration, we then perform a linear interpolation between the nearest two measurements. Due to the dynamic criteria of integration time $t_{\rm int}$ (see Section.~\ref{subsec:vis}), one should expect longer $t_{\rm int}$ and larger $N_{\rm eff}$ at lower frequencies where the larger changes of baselines $|\delta \textbf{r}_{ij}|$ can be tolerated, and on shorter baselines where the relative velocities are lower.
This effective error model is denoted as the ``Real'' model. 

\subsection{The instrumental phase error}
The instrument phase of the receiver can be calibrated, but there could be residual error in this process. Such instrumental phase errors $\Delta\phi^{\rm instru}$ are induced when cross-correlations are performed on the raw voltage data received from the antennas. Therefore, they are modeled as being directly added to the phase of visibility functions, i.e.,
\begin{equation}
    V^{\rm err}_{ij}=V^{\rm true}_{ij}{\exp}({-{\mathrm i}\,\Delta\phi^{\rm instru}_{ij}})\,.
\end{equation}
Here $V^{\rm true}_{ij}$ is the true visibility, $V^{\rm err}_{ij}$ is the data we will have, and $\Delta\phi^{\rm instru}$ is unknown in the reconstruction process.

If all the instruments are well tested and calibrated, $\Delta\phi_{\rm instru}$ should be randomly distributed with a zero mean. In this case, we model them as independent Gaussian random numbers with a HWHM of $\Delta\phi_0$. However, we will also consider the possibility that there might be constant bias terms due to calibration bias or unpredictable reasons. 
Such constant errors are hard to remove via post-processing of visibility functions on the ground.
In this case, we model them as a constant value $\Delta\phi_0$, and assess its impact. 

In principle, the thermal cycling in orbit could induce time-varying phase drifts. However, phase consistency in the DSL mission is robustly maintained by calibrations \citep{chen2023}, which correct for environmental effects, including thermal fluctuations. Therefore, time-varying phase drifts are negligible and not considered in our analysis.

As a brief summary, in Table~\ref{tab:suppresion_gaussian}, we list all the models of phase errors discussed in this paper, and the theoretical estimation of the relative error in terms of angular power spectrum (to be discussed in the next section).

\begin{table}
\centering
\caption{The theoretical estimation of $C^{error}_\ell/C^0_\ell$ for various phase error models in the scheme of all-sky imaging.}
\begin{tabular}{c|c|c|c}
\hline\hline
     Freq [MHz]& NSIDE & Error Model & $C^{error}_\ell/C^0_\ell$\\
     \hline
     30 & 16 & Constant $12^\circ$ & 0.9568\\
     30 & 16 & Constant $25^\circ$ & 0.8214\\
     30 & 16 & Constant $50^\circ$ & 0.4132\\
     30 & 16 & Random $12^\circ$ & 0.9689\\
     30 & 16 & Random $25^\circ$ & 0.8717\\
     30 & 16 & Random $50^\circ$ & 0.5774\\
     3 & 16 & Random $50^\circ$ & 0.5774\\
     10 & 16 & Random $50^\circ$ & 0.5774\\
     30 & 32 & Random $50^\circ$ & 0.5774\\
     30 & 64 & Random $50^\circ$ & 0.5774\\
     30 & 16 & Single 1\,m &0.5646 \\ 
     30 & 16 & Real 1\,m & 0.9898 \\ 
     30 & 16 & Real 3\,m & 0.9133 \\ 
     30 & 16 & Real 5\,m & 0.7828 \\ 
     3 & 16 & Real 1\,m &0.9999 \\
     10 & 16 & Real 1\,m &0.9996 \\
     30 & 32 & Real 1\,m &0.9898 \\
     30 & 64 & Real 1\,m &0.9898 \\
     30 & 16 & Random 1.1\,ns& 0.9408\\ 
     30 & 16 & Random 3.3\,ns& 0.5774\\ 
     30 & 16 & Random 5.5\,ns& 0.2175\\ 
     3 & 16 & Random 3.3\,ns& 0.9945\\ 
     10 & 16 & Random 3.3\,ns& 0.9408\\ 
     30 & 32 & Random 3.3\,ns& 0.5774\\ 
     30 & 64 & Random 3.3\,ns& 0.5774\\ 
\hline
\end{tabular}
\label{tab:suppresion_gaussian}
\end{table}

\section{Results}
\label{sec:results}
In this section, we present the impacts of phase errors. For illustrative purposes, we present a simple example in Section~\ref{subsec:example}, where phase errors are added to the Fourier modes of a 2D image, to observe the general impacts of the phase error. We then discuss the impacts on all-sky imaging in Section~\ref{subsec:all-sky} and on the imaging of a patch of sky in Section~\ref{subsec:patchy-sky}.

\subsection{A Simple Example}
\label{subsec:example}
\begin{figure*}
    \centering
    \includegraphics[width=0.64\columnwidth]{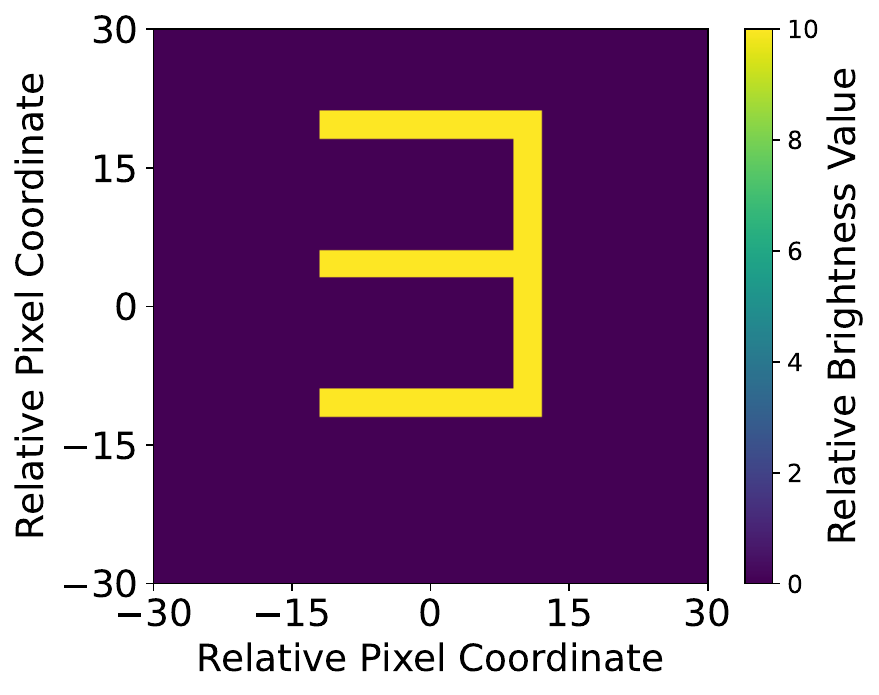}
    \includegraphics[width=0.64\columnwidth]{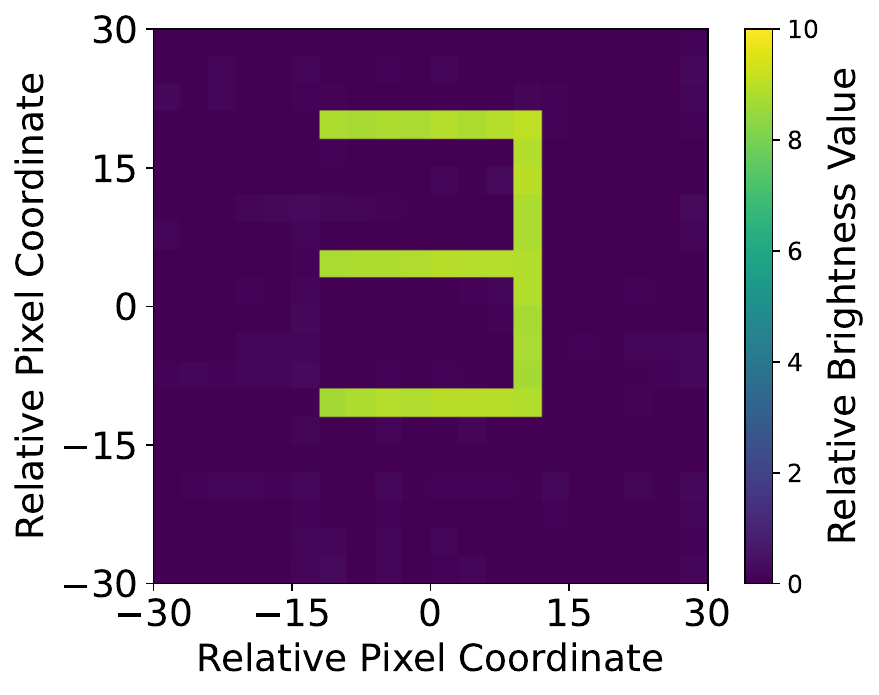}
    \includegraphics[width=0.64\columnwidth]{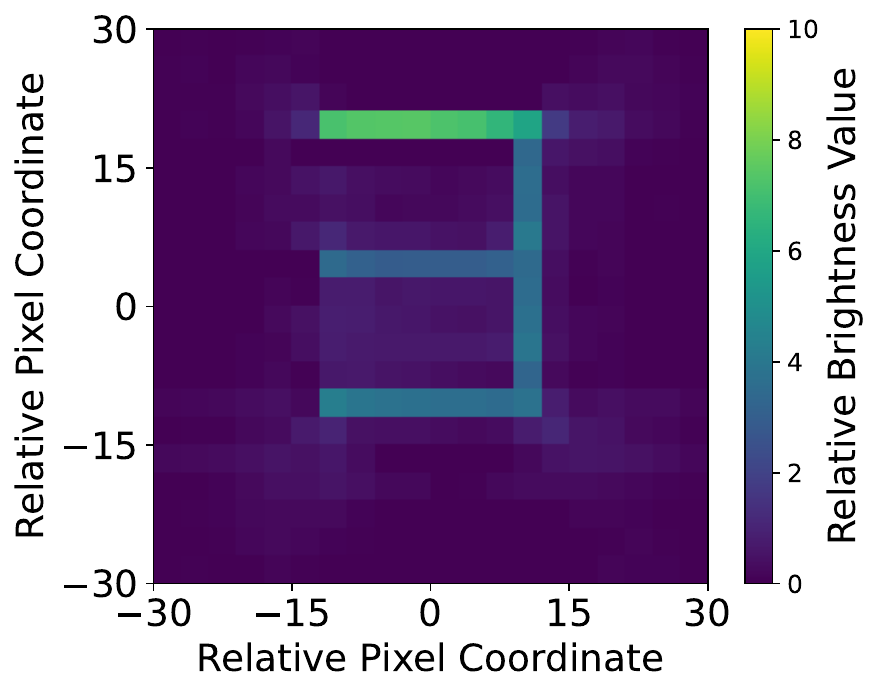}
    \caption{Illustrative examples with random phase errors. We plot the reconstructed images without phase errors (Left), with direction-independent random phase errors (Middle), and with direction-dependent random phase errors (Right). The x- and y-axes show relative pixel coordinate, and the color bar indicates relative brightness value. These are representative illustrative examples, and no absolute physical units are assigned.}
    \label{fig:illustrative}
\end{figure*}

We consider a simple 2D flat image, e.g. an artificial image in the shape of an ``$\exists$'' sign. In an idealized interferometric imaging, we would obtain all its Fourier components as visibilities, and reconstruct the image by an inverse Fourier transform. We can add an error to the phase of each Fourier component and see how this affects the image. This is shown in the three panels of Figure~\ref{fig:illustrative}, which shows the reconstruction result without phase errors (left), with isotropic random phase errors (center), and with random phase errors (right) whose distribution depends on the direction of ``baseline vectors''---the Fourier mode vectors. 

In the simple case of a flat 2D image, a phase difference of a Fourier component would make a translational shift of that Fourier component in the image. Thus, with an error added to the phase of each Fourier component, that component is moved away from its original position, spreading some brightness to the background of the image. But if the phase errors are randomly distributed with zero mean, the original image would still remain with a somewhat decreased brightness, and the diffuse background is enhanced with some random noise, as shown in the center panel of Figure~\ref{fig:illustrative}. On the other hand, if the phase errors are not random, e.g. if they depend on the direction of the baseline vector, then this effect is much more apparent, as shown in the right panel of Figure~\ref{fig:illustrative}. This example indicates that generally, phase errors in radio interferometric imaging can result in two effects: signal degradation and image distortion. 

We can make some theoretical predictions. If we neglect confusion noise and thermal noise, we can formally rewrite $\mathbf{s}^{\rm dirty}$ in Section~\ref{sec:algorithm} as follows:
\begin{eqnarray}
\label{eqn:dirtymap}
    s^{\rm dir}_\beta &=& \sum_{\gamma} \tilde{D}_{\beta\gamma}s_\gamma
\end{eqnarray}
where $\beta$ and $\gamma$ denote the index of the sky direction, and $\tilde{D}_{\beta\gamma}={\rm Re}(\sum_\alpha B_{\alpha\beta}^*B_{\alpha\gamma})$ is the real part of the dirty beam. If there are some phase errors $\phi_{\alpha\beta}$ on $\mathbf{r}_\alpha$ which contribute to the sky direction $\hat{\mathbf{n}}_\beta$, the dirty map with phase error $\mathfrak{s}^{\rm dir}$ is given by,
\begin{eqnarray}
    \mathfrak{s}^{\rm dir}_\beta &=& {\rm Re}[\sum_{\alpha\gamma} B_{\alpha\beta}^*B_{\alpha\gamma}\exp(\mathrm i\phi_{\alpha\beta})s_\gamma]\nonumber\\
    &=& \sum_{\gamma}(\tilde{D}_{\beta\gamma}\overline{\cos\phi_{\alpha\beta}}-\tilde{D}^{\prime}_{\beta\gamma}\overline{\sin\phi_{\alpha\beta}}) s_\gamma
\end{eqnarray}
where we define $\tilde{D}^\prime_{\beta\gamma}={\rm Im}(\sum_\alpha B_{\alpha\beta}^*B_{\alpha\gamma})$, which is the imaginary part of the dirty beam, $\overline{\cos\phi_{\alpha\beta}}$ and $\overline{\sin\phi_{\alpha\beta}}$ are averaged over baselines. The $\tilde{D}^\prime_{\beta\gamma}$ term should vanish at large scales because 
contributions from pixels at opposite sky directions cancel out with each other. 

For the random phase errors, 
we can use the expectation of the overall dirty beam to approximate the effective dirty beam for measured visibility data, i.e.,
\begin{eqnarray}
\label{eqn:random_dirty_beam}
    \widehat{D}^{\rm rand}_{\beta\gamma}\approx\langle\tilde{D}_{\beta\gamma}\rangle
    = \tilde{D}_{\beta\gamma}\langle{\mathrm{cos}}\phi\rangle_\beta-\tilde{D}^{\prime}_{\beta\gamma}\langle{\mathrm{sin}}\phi\rangle_\beta
\end{eqnarray}
where $\langle{\mathrm{cos}}\phi\rangle_\beta$ and $\langle{\mathrm{sin}}\phi\rangle_\beta$ are the expectations of $\overline{\cos\phi_{\alpha\beta}}$ and $\overline{\sin\phi_{\alpha\beta}}$, respectively. If $\phi_{\alpha\beta}$ is uncorrelated with $\hat{\mathbf{n}}_\beta$, $\langle{\mathrm{sin}}\phi\rangle$ should be zero and $\widehat{D}^{\rm rand}_{\beta\gamma}$ should decrease with a factor of $\langle{\mathrm{cos}}\phi\rangle$, resulting in a uniform degradation of $\mathfrak{s}^{\rm dir}_\beta$. On the other hand, if the distribution of $\phi_{\alpha\beta}$ depends on $\hat{\mathbf{n}}_\beta$, then $\langle{\mathrm{cos}}\phi\rangle_\beta$ may depend on $\hat{\mathbf{n}}_\beta$ and the reconstructed image will be distorted.

If $\phi_{\alpha\beta}$ is a constant value $\Delta\phi_0$ over baselines, which is a strong systematics, the overall dirty beam should be given by
\begin{eqnarray}
\label{eqn:const_dirty_beam}
    \widehat{D}^{\rm const}_{\beta\gamma} = \tilde{D}_{\beta\gamma}\cos\Delta\phi_0-\tilde{D}^{\prime}_{\beta\gamma}\sin\Delta\phi_0\,.
\end{eqnarray}
Therefore, at large scales, without the contribution of $\tilde{D}^{\prime}_{\beta\gamma}$, $\tilde{D}^{\rm const}_{\beta\gamma}$ and $\mathfrak{s}^{\rm dir}_\beta$ should both decrease with a factor of $\cos\Delta\phi_0$. While at small scales, the survived $\tilde{D}^\prime_{\beta\gamma}$ can also result in image distortion. 

In the following subsections, we explore how the impacts of various kinds of phase errors quantitatively depend on the error model, magnitude, frequency, and spatial resolution based on the simulated results.


\begin{figure*}[htbp]
    \centering
    \includegraphics[height=0.8\columnwidth]{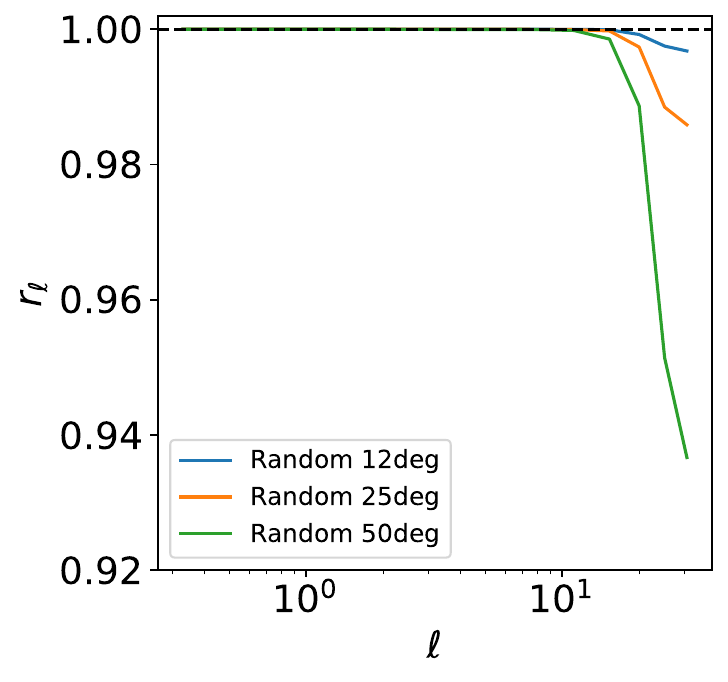}
    \includegraphics[height=0.8\columnwidth]{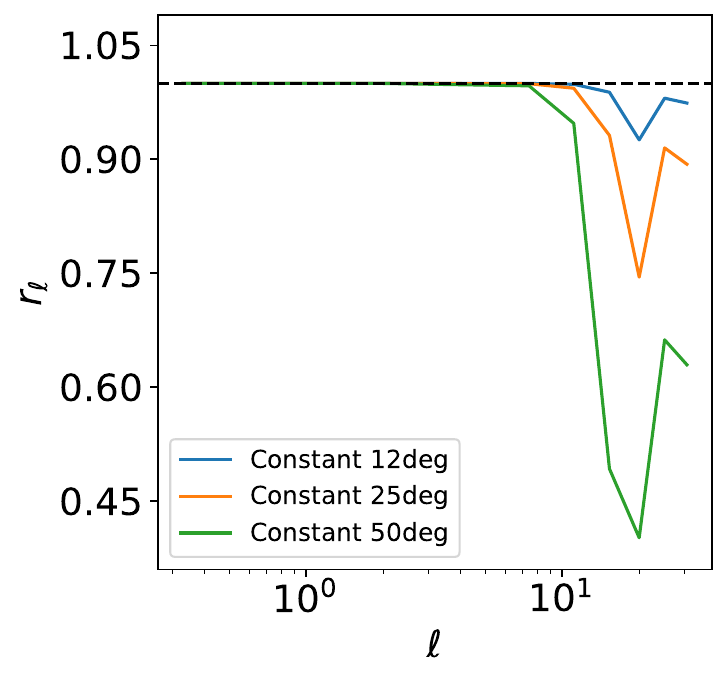}
    \caption{The correlation coefficient $r_\ell$ between the reconstructed maps with and without phase errors as a function of $\ell$ for various instrumental phase errors. We model the error as constant deviations of $\Delta\phi_0=12^\circ$ (blue), $25^\circ$ (orange), and $50^\circ$ (green) in the right panel, while Gaussian random values of $\Delta\phi_0=12^\circ$ (blue), $25^\circ$ (orange), and $50^\circ$ (green) in the left panel, respectively.}
    \label{fig:all-sky instru rl}
\end{figure*}

\subsection{Impacts on all-sky imaging}
\label{subsec:all-sky}
In the scenario of all-sky imaging, we focus on how well one can reconstruct the large-scale structure of the ultra-long wavelength sky at various scales with different sources of phase errors, and ignore the thermal noise first. 

We will use only the visibility data measured with projected baselines shorter than 1.2\,km so that the fractional error in baselines is fixed. This is four times the Nyquist limit at $f=30$\,MHz and NSIDE=64. which should be sufficient for the cases
listed in Table~\ref{tab:suppresion_gaussian}. Longer baselines do not significantly improve the quality of all-sky imaging at a resolution lower than that defined by NSIDE=64 \citepalias{2025arXiv251118494D}.
Here, we only use the diffuse component of the input sky map, since the point sources would not affect large-scale structures in all-sky imaging. 

\begin{figure*}[htbp]
    \centering
    \includegraphics[height=0.8\columnwidth]{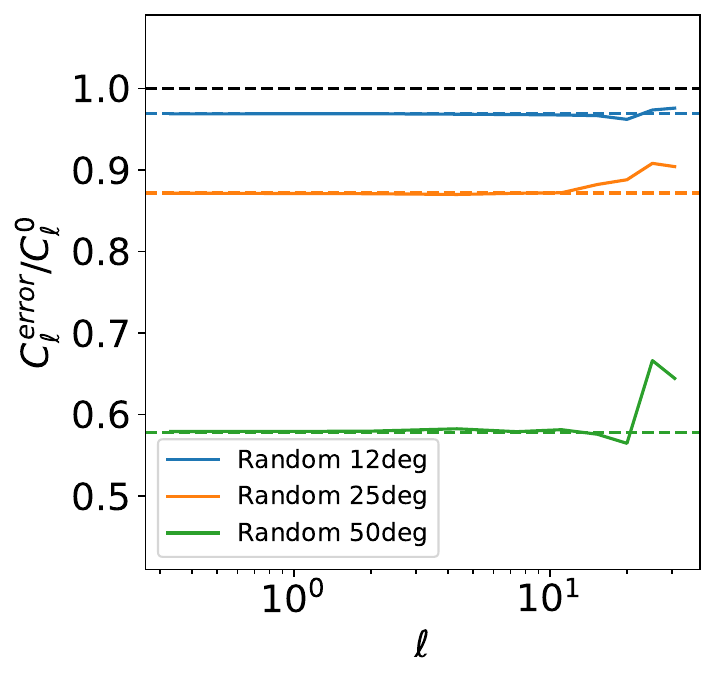}
    \includegraphics[height=0.8\columnwidth]{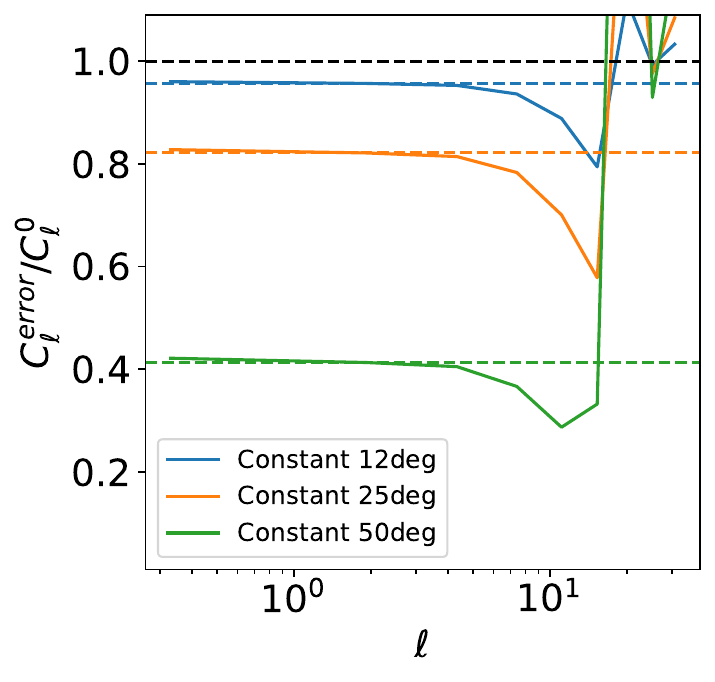}
    \caption{The ratio $C_\ell^{error} /C_\ell^{0}$ as a function of $\ell$ for various instrumental phase errors. We model the error as constant deviations of $\Delta\phi_0=12^\circ$ (blue), $25^\circ$ (orange), and $50^\circ$ (green) in the right panel, while Gaussian random values of $\Delta\phi_0=12^\circ$ (blue), $25^\circ$ (orange), and $50^\circ$ (green) in the left panel, respectively. The solid lines are the simulated results, while the colored dashed lines are the theoretical estimations.}
    \label{fig:all-sky instru}
\end{figure*}

In order to quantify the impacts of the instrumental phase error, we first compute the correlation coefficient between the reconstructed maps and the input map, 
\begin{eqnarray}
    r_\ell = \frac{C^X_\ell}{\sqrt{C^{error}_\ell C^0_\ell}}\,,
\end{eqnarray}
where $C^X_\ell$ is the cross angular power spectrum between the reconstructed maps with phase error and the input map, $C^{error}_\ell$ is the angular power spectrum of the reconstructed maps with phase error, and $C^0_\ell$ is the angular power spectrum of the original input map. We plot $r_\ell$ as a function of $\ell$ in Figure~\ref{fig:all-sky instru rl}. For the constant phase errors, $r_\ell$ decreases significantly at $\ell>10$, showing an image distortion on smaller scales. In contrast, for random instrumental phase errors and other phase error sources, the image fidelity of reconstructed sky maps is well preserved, $r_\ell$ remains very close to 1, only drops very slight at  $\ell>10$. Similar to the simple example of Section~\ref{subsec:example}, the main effect is a degradation of the reconstructed image. 

\begin{figure*}
    \centering
    \includegraphics[height=0.8\columnwidth]{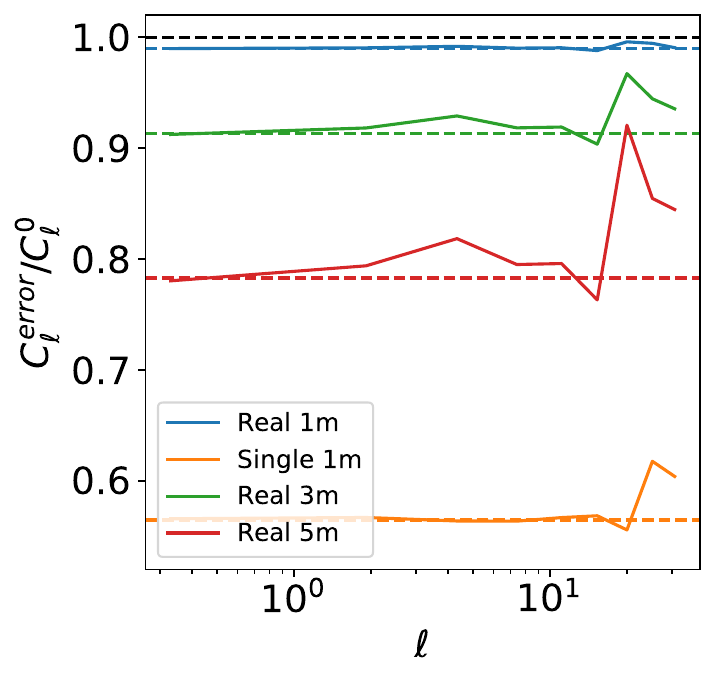}
    \includegraphics[height=0.8\columnwidth]{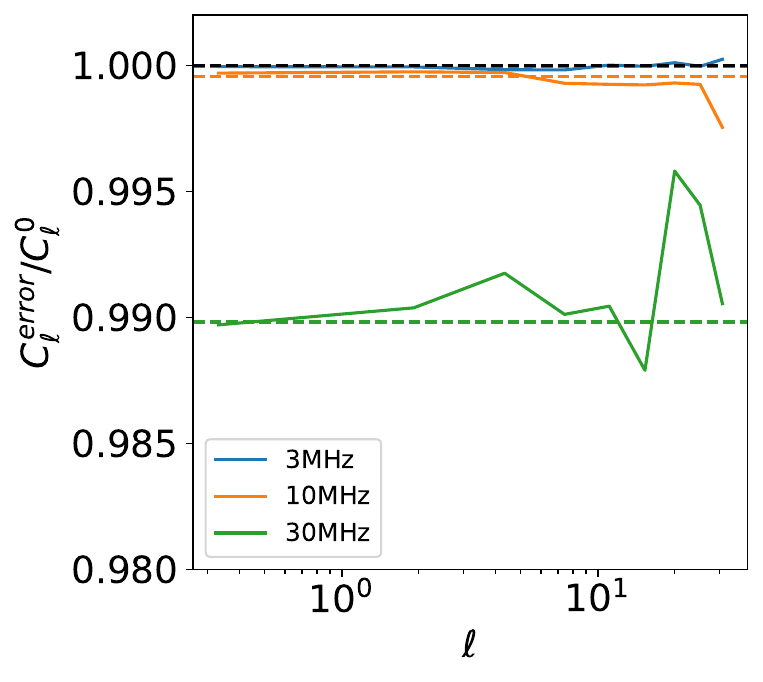}
    \caption{The ratio $C_\ell^{error} /C_\ell^{0}$ as a function of $\ell$ for geometric phase errors. We vary the error models of $\Delta r_0=1$\,m (blue for ``Real'' model and orange for ``Single'' model), $3$\,m (green), and $5$\,m (red) at 30\,MHz in the left panel, while vary the frequency of $f=3$\,MHz (blue), $10$\,MHz (orange), and $30$\,MHz (green) with the fixed ``Real'' error model of 1\,m in the right. The solid lines are the simulated results, while the colored dashed lines are the theoretical estimations.}
    \label{fig:all-sky geo}
\end{figure*}

To quantify this signal degradation, we plot the ratio of $C_\ell^{error}$ to $C_\ell^{0}$ at $f=30$\,MHz and NSIDE=16, as a function of $\ell$ in Figure~\ref{fig:all-sky instru}. We model the error as a constant deviation with different values in the right panel and random Gaussian variables with different HWHMs in the left panel, respectively. 
We find uniform suppression factors on $C^0_\ell$ on large scales in both cases, as predicted in Section~\ref{subsec:example}. Constant phase errors result in more significant suppressions in large-scale power than random phase errors. 
The uniform suppressions for constant phase errors break down at $\ell\sim10$.
We also use Equations~(\ref{eqn:random_dirty_beam}) and (\ref{eqn:const_dirty_beam}) to theoretically estimate the suppression factors and list them in Table~\ref{tab:suppresion_gaussian}. In principle, the ratio $C^{error}_\ell/C_\ell^0$ should be $\langle{\mathrm{cos}}\phi\rangle^2$ for random phase errors, and $\cos^2\Delta\phi_0$ for constant phase errors. The simulated results are consistent with the theoretical estimation (colored dashed lines in Figure~\ref{fig:all-sky instru}). Additionally, we find that the suppression factors do not depend on the spatial resolution or frequency for the instrumental phase errors.

\begin{figure*}
    \centering
    \includegraphics[height=0.8\columnwidth]{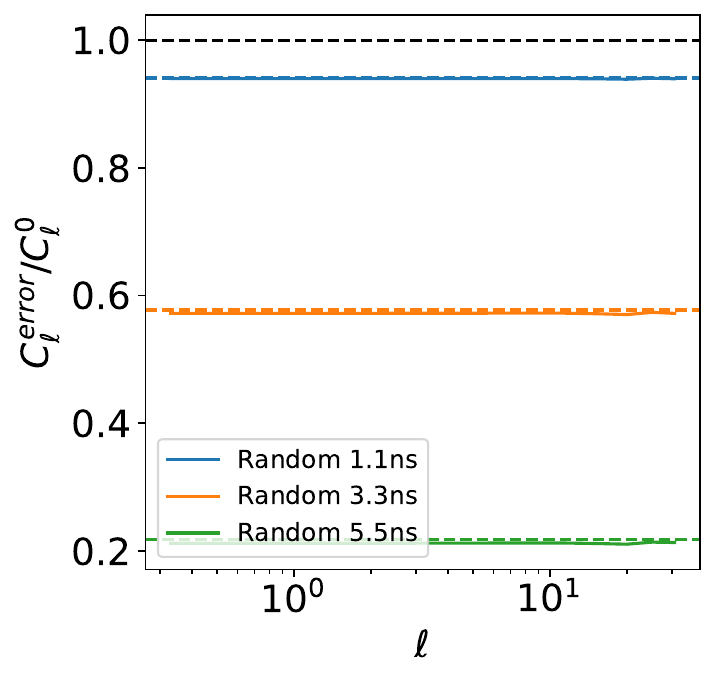}
    \includegraphics[height=0.8\columnwidth]{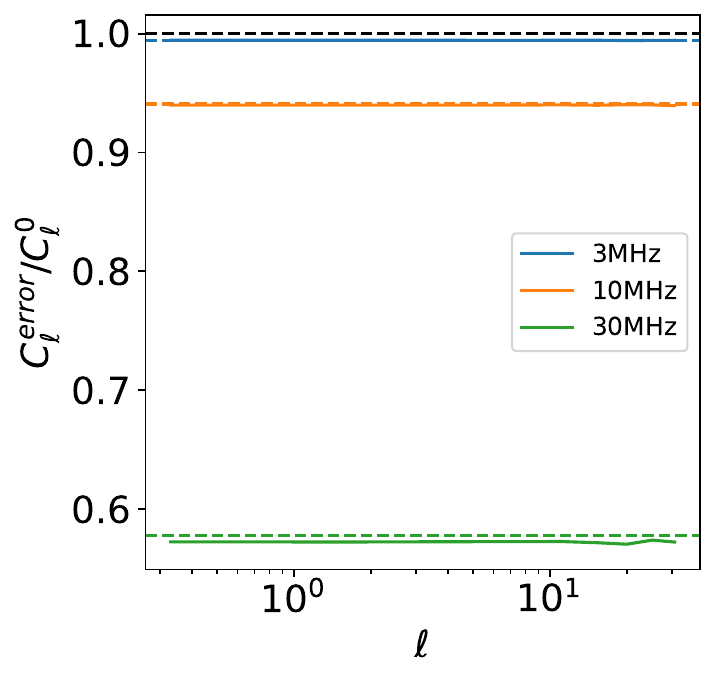}
    \caption{The ratio $C_\ell^{error}/C_\ell^{0}$ as a function of $\ell$ for temporal phase errors. We vary the error magnitude of $\Delta t_0=1.1$\,ns (blue), $3.3$\,ns (orange), and $5.5$\,ns (green) at 30\,MHz in the left panel, while vary the frequency of $f=3$\,MHz (blue), $10$\,MHz (orange), and $30$\,MHz (green) with the fixed magnitude of $\Delta t_0=3.3$\,ns in the right. The solid lines are the simulated results while the colored dashed lines are the theoretical estimations.}
    \label{fig:all-sky time}
\end{figure*}

For the geometric phase errors, similarly, the distortion is very small, the main effect is the degradation of the power spectrum. We plot $C_\ell^{error}/C_\ell^0$ as a function of $\ell$ at NSIDE=16 in Figure~\ref{fig:all-sky geo}. In the left panel, we vary the error model at 30\,MHz, while vary the frequency with the ``Real'' error model of 1\,m in the right panel. 
We find uniform suppression factors on large scales similar to the case of random instrumental phase errors.
The ``Single'' model results in a stronger degradation than the ``Real'' model with fixed $\Delta r_0$. This is consistent with our expectation because effective baseline errors should be reduced after multiple measurements. We use the distribution of effective baseline errors to estimate $C_\ell^{error}/C_\ell^0$ of all cases (see Appendix~\ref{appendix_b} for details).
The simulation results also match the theoretical estimates. Suppression factors do not depend on spatial resolution. However, we do notice that $C_\ell^{error}/C_\ell^0$ at high frequencies is more affected. There are two reasons for the frequency dependence. First, the effective baseline errors 
at higher frequencies are larger (see Figure~\ref{fig:bl_error}), which results in stronger suppression. Second, the geometric phase errors are also proportional to frequency explicitly since $\Delta\phi_{\rm geo}$ should scale with $|\Delta\mathbf{r}_{ij}|/\lambda$. Even with the same effective baseline errors, the suppression should still be stronger at high frequencies.

Again, for the temporal phase error, the distortion is very small, and the effect is mainly in the degradation of the angular power spectrum of the signal. 
We also plot $C_{\ell}^{error}/C_\ell^0$ at NSIDE=16 with temporal phase errors in Figure~\ref{fig:all-sky time}. In the left panel of Figure~\ref{fig:all-sky time}, we vary the error magnitude at 30\,MHz, while vary the frequency with the fixed magnitude of $\Delta t=3.3$\,ns in the right panel. We also list the theoretical estimates in Table~\ref{tab:suppresion_gaussian}. Again, we find uniform suppression factors consistent with the theoretical estimates. 
They are not affected by spatial resolutions but get stronger at higher frequencies because $\Delta\phi^{\rm temporal}$ increases with frequency explicitly according to Equation (\ref{eqn:time_err}). Since the overall impacts of temporal phase errors have been averaged within each $t_{\rm int}$ before being averaged over baselines, the suppression factors are more uniform at different scales.

We also notice that at the same time of reduced angular power on large scales, there is an excess power at small scales with the phase errors. In the case of constant phase errors, the large excess power comes from image distortion due to the second term in Equation~(\ref{eqn:const_dirty_beam}). In the case of random phase errors, the ``sample variance'' in quasi-redundant baselines (in one $uvw$ cell) can generate fluctuations on visibilities, which contribute excess power on small angular scales. 

We can set up requirements on phase error budgets based on the criteria that the degradation on $C_\ell^0$ can be comparable to the thermal noise level. The thermal noise angular power spectrum at the maximum $\ell\sim47$ corresponding to NSIDE=16 is roughly 5\% of the intrinsic $C_\ell^0$ at 30MHz. So, a tolerance of a 5\% decrease in the angular power spectrum is deemed reasonable across frequencies and error models.
Therefore, 
we require a threshold of $\Delta\phi_0\sim 12^\circ$ on the constant component of instrumental phase error, or $\Delta\phi_0\sim 12^\circ$ on the random component of instrumental phase error, or $\Delta t_0\sim 1.1$ ns on the temporal phase error. With post-processing of multiple baseline measurements, $\Delta r_0\sim 1$m can also meet the requirement. For quick reference, we summarize error tolerances in Table~\ref{tab:threshold_gaussian}.

\begin{table}
\centering
\caption{The error tolerance for various phase error models in the scheme of all-sky imaging.}
\begin{tabular}{c|c}
\hline\hline
     Error Model & Error Tolerance\\
     \hline
     Instrumental: Constant & $12^\circ$\\
     Instrumental: Random & $12^\circ$\\
     Geometric: Real  & 1\,m \\ 
     Temporal: Random & 1.1\,ns\\ 
\hline
\end{tabular}
\label{tab:threshold_gaussian}
\end{table}

%




\subsection{Impacts on patchy-sky imaging}
\label{subsec:patchy-sky}
In the scenario of patchy-sky imaging, we focus on point source detections. 
In practice, one should apply some source identification algorithm to search for point sources and estimate their signal-to-noise ratios (S/N). However, here we focus on the impact of phase errors but skip the source search in this work, and assume that we will be able to identify and locate the sources accurately, at least for the bright sources.

\begin{figure}
    \centering
    \includegraphics[height=0.8\columnwidth]{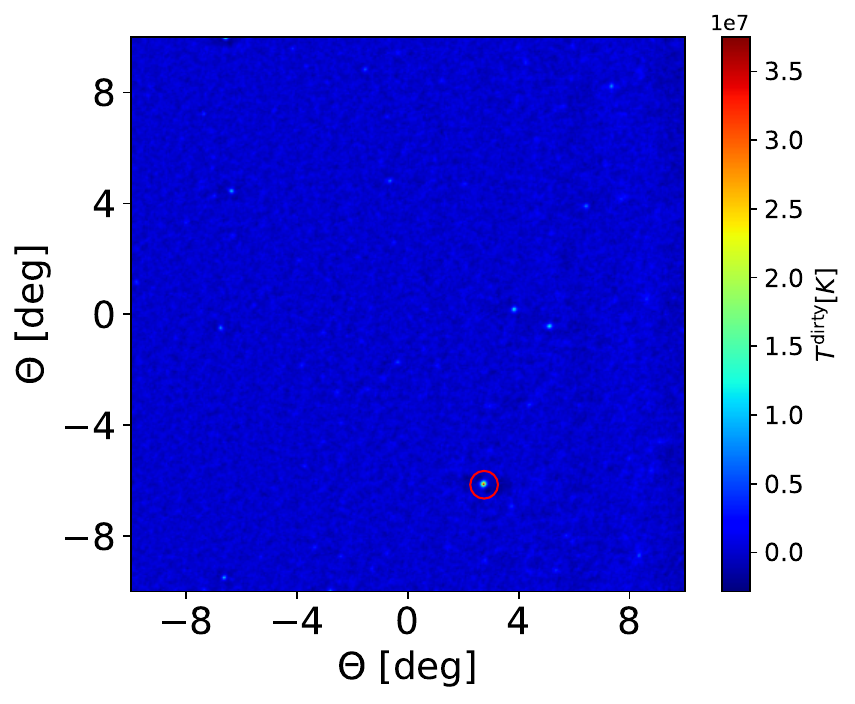}
    \caption{The reconstructed dirty map of the $20^\circ\times20^\circ$ patchy-sky area centered at RA=75$^\circ$, DEC=-45$^\circ$. We circle the brightest point source here, whose S/N we choose to forecast.}
    \label{fig:patchy-sky imaging}
\end{figure}

To separate the unknown systematics from the point source identification criteria, we perform the beam-forming imaging algorithm on a 20$^\circ\times$20$^\circ$ patchy-sky area centered at (RA=75$^\circ$, DEC=-45$^\circ$).
We use the visibility data measured with baselines up to 100\,km so that we can capture the information at the smallest scale. In order to lower the confusion noise from the diffuse background when we estimate the S/N of point sources, we also subtract the diffuse background contribution by excluding visibility data measured with baselines whose projected lengths are smaller than $\sim 9$\,km. We find that most powers from the diffuse background can be subtracted with this cutoff. We analyze the S/N of the brightest point source in the field, which is circled in Figure~\ref{fig:patchy-sky imaging}, to quantify the impact of the phase errors. The flux of this point source is 831.96 Jy at 3 MHz, and we assume a spectral index of 0.8.

In principle, a point source shall be identified if its flux density is higher than the diffuse background, with the significance determined by the thermal noise. Therefore, the S/N of an identified point source should be
\begin{equation}
    {\rm S/N} = \frac{T_{\rm source}-T_{\rm background}}{\sigma_{\rm patchy}}\,,
\end{equation}
where $\sigma_{\rm patchy}$ is the standard deviation of brightness temperature in the patchy-sky area, $T_{\rm background}$ is the mean brightness temperature of the diffuse component near the source. 
Since we subtract the diffuse background contribution, $T_{\rm background}$ is nearly zero for the sky patch in Figure~\ref{fig:patchy-sky imaging}. 

\begin{figure*}
    \centering
    \includegraphics[height=0.64\columnwidth]{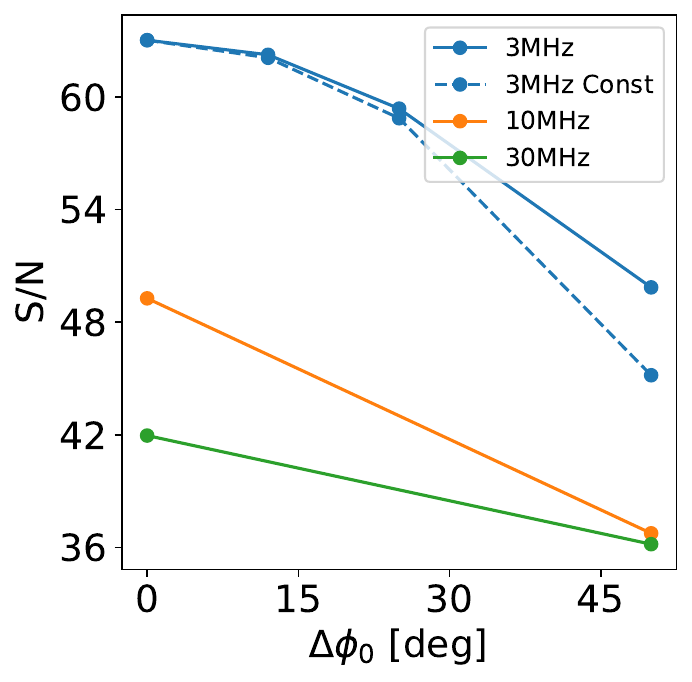}
    \includegraphics[height=0.64\columnwidth]{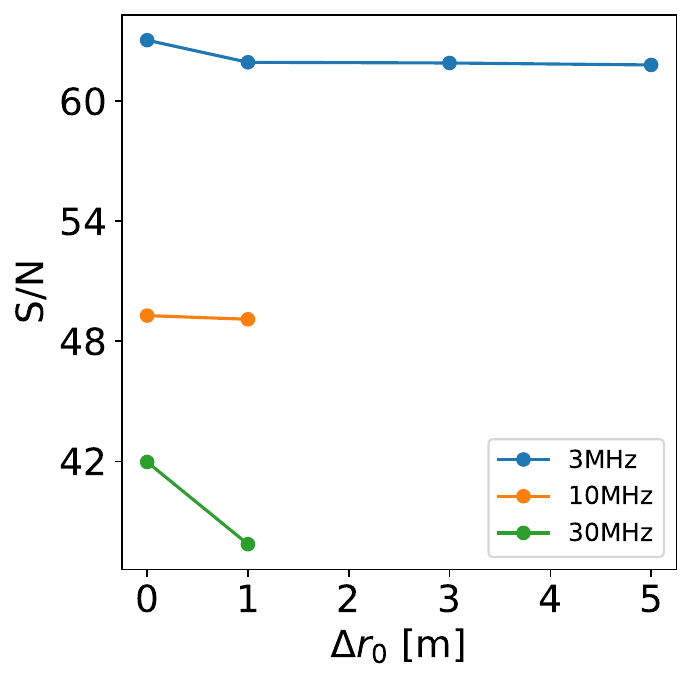}
    \includegraphics[height=0.64\columnwidth]{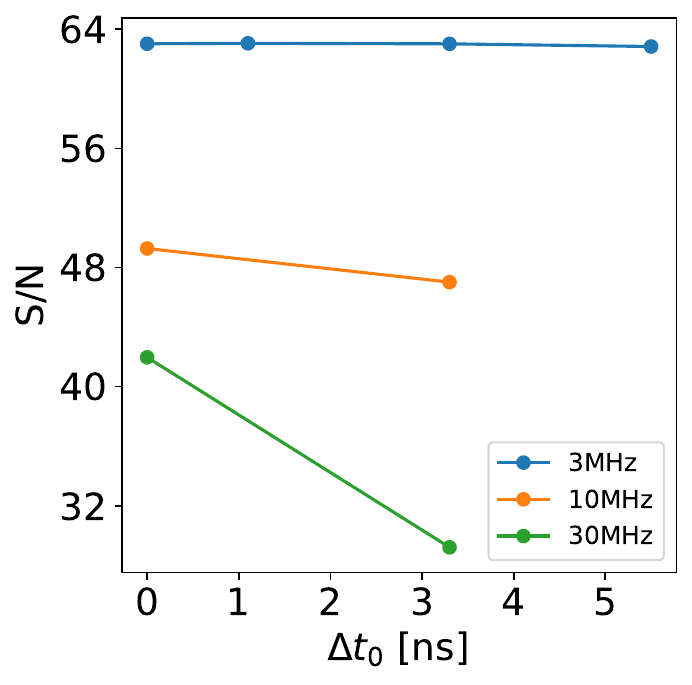}
    \caption{The S/N of the brightest point source with (from left to right) instrumental, geometric, and temporal phase errors as a function of error magnitude. In the left panel, the solid lines show the S/N values with random instrumental phase errors, and the dashed line shows the results with constant instrumental phase errors for 3 MHz. In all cases, we vary the frequency of $f=3$\,MHz (blue), $10$\,MHz (orange), and $30$\,MHz (green) along with spatial resolution simultaneously to make sure that information at the smallest scales can be captured with 100\,km baselines. 
    }
    \label{fig:snr_patchy}
\end{figure*}

We plot the S/N of the brightest point source with instrumental, geometric, and temporal phase errors, respectively in Figure~\ref{fig:snr_patchy}, all as a function of the error magnitudes. We vary the frequency and spatial resolution simultaneously to make sure that information at the smallest scales can be captured with 100\,km baselines. We find that random geometric and temporal phase errors do not significantly change the S/N at low frequencies. 
They moderately reduce the S/N at $f=30$\,MHz 
with $\Delta r_0=1$ m, and 
with $\Delta t_0=3.3$ ns, respectively. This is consistent with the impacts on all-sky imaging that imaging at high frequencies is affected more severely. We also observe noticeable decreases in S/N with instrumental phase errors at all frequencies. The S/N values have a fractional decrease of $5-10\%$ with random instrumental phase errors of $\Delta\phi_0=50^\circ$. The impact of a constant instrumental phase error is the worst. At $f=3$\,MHz, a constant instrumental phase error of $\Delta\phi_0=50^\circ$ can reduce the S/N with a fraction over $25\%$.

In principle, we expect $T_{\rm source}$ to degrade with phase errors by the same suppression factors as in 
Sec~{\ref{subsec:all-sky}}, except for geometric phase errors, in which case the effective errors of longer baselines are larger. However, the impacts of phase errors on the S/N of point sources are different because $\sigma_{\rm patchy}$ can also be degraded.
If we assume that the confusion noise and thermal noise are roughly independent, $\sigma_{\rm patchy}$ can be decomposed as
\begin{equation}
    \sigma^2_{\rm patchy} = \sigma^2_{\rm confusion}+\sigma^2_{\rm thermal}\,,
\end{equation}

where $\sigma_{\rm confusion}$ is the confusion noise level and $\sigma_{\rm thermal}$ is the thermal noise level. 
Since the real and imaginary parts of thermal noise in visibility are incoherent, we expect no significant suppression on $\sigma_{\rm thermal}$ due to the various phase errors. $\sigma_{\rm confusion}$ can be regarded as the reconstruction error of the diffuse background, which should be suppressed in a similar way to $T_{\rm source}$. The overall $\sigma_{\rm patchy}$ can also be degraded. Therefore, the impacts of instrumental and temporal phase errors on S/N of point sources are weaker than their impacts on the all-sky angular power spectrum.
For the geometric phase errors, since we use visibility data of longer baselines whose measurement counts are fewer and the effective errors are larger, the impact on the S/N of point sources is stronger than in the scheme of all-sky imaging, due to a more severe suppression on the signal, but still tolerable. Note that the discussion above is only valid for point sources. With respect to extended sources, we expect the confusion noise to be much less dominant. The impact of phase errors on extended sources should be more like the all-sky imaging scheme.

Therefore, we suggest that the requirements on instrumental and temporal phase errors should be established by the quality requirement of all-sky imaging for future space interferometers, while the requirements on geometric phase errors should be determined by the quality requirement of patchy-sky or small-scale imaging.
\section{Conclusion}
\label{sec:conclusion}
In this work, we have investigated the impacts of phase errors in synthesis imaging with a lunar orbit array. We adopt the configuration of the DSL mission and simulate the visibility measurements with Moon blockage. We use the pixel-averaging linear brute-force map-making method to perform all-sky imaging and the beam-forming algorithm to perform patchy-sky imaging. We model instrumental phase errors, geometric phase errors, and temporal phase errors separately and vary the error magnitudes, frequencies, and spatial resolutions to reconstruct dirty sky maps for analysis. Our main findings and conclusions are summarized as follows.

\begin{itemize}
    \item In the scheme of all-sky imaging, we find uniform suppression factors on the angular power spectrum on large scales for various sources of phase errors. The uniform suppressions for constant instrumental phase errors break down at a scale larger than the angular resolution that the longest baselines can probe. 
    Multiple baseline determinations can reduce effective baseline errors and alleviate the suppression factor.
    The suppressions due to geometric and temporal phase errors are both weaker at lower frequencies, with the former because of smaller effective errors and the latter because of the scaling relations between phase errors and measurement errors. 
    \item In the scheme of patchy-sky imaging, we find that the S/N of point source detection is more severely affected at higher frequencies or with constant instrumental phase errors. The impact of geometric phase error is stronger because the effective errors of longer baselines are larger. In other scenarios, the decreases of the S/N values are weaker because the overall noises are also suppressed.
    \item The requirement from the S/N of point sources should be less strict than that from the suppression of the all-sky angular power spectrum. With a tolerance of a 5\% decrease in the angular power spectrum, a threshold of $\Delta\phi_0\sim 12^\circ$ on the constant component of instrumental phase error, or $\Delta\phi_0\sim 12^\circ$ on the random component of instrumental phase error, or $\Delta t_0\sim 1.1$ ns on the temporal phase error should be required. With post-processing of multiple baseline measurements, $\Delta r_0\sim 1$\,m can also meet the requirement.  
\end{itemize}

These analyses present quantitative impacts of the various instrument-induced phase errors on the imaging quality. The resultant error budgets allowed for a reasonable imaging quality, represent the basic requirements for the instrumental design of the DSL, as well as for any future space interferometers. These results are consistent with other recent works on space interferometry
errors if rescaled to similar criteria \citep{2013aero.confE.120R,2026MNRAS.546ag116G}. For future static lunar-surface arrays such as LARAF \citep{2024arXiv240316409C}, on the engineering level, there is still a possibility that they will measure baseline vectors and do clock synchronization after construction. Although they may not use the same approach as orbital arrays, the impacts due to phase errors should be similar. Therefore, the phase error analysis framework established in this paper can benefit both orbital and surface-based lunar interferometric missions.


\section*{Acknowledgments}
We thank Li Deng for helpful discussions. This work was supported by the National Key R\&D Program of China No. 2022YFF0504300, China's Space Origins Exploration Program Nos. GJ11010401 and GJ11010405, the NSFC International (Regional) Cooperation and Exchange Project No. 12361141814, and by the Specialized Research Fund for State Key Laboratory of Radio Astronomy and Technology.

%

\vspace{5mm}




\appendix

\section{The systematics from deconvolution}
\label{appendix_a}
\begin{figure}
    \centering
    \includegraphics[height=0.8\columnwidth]{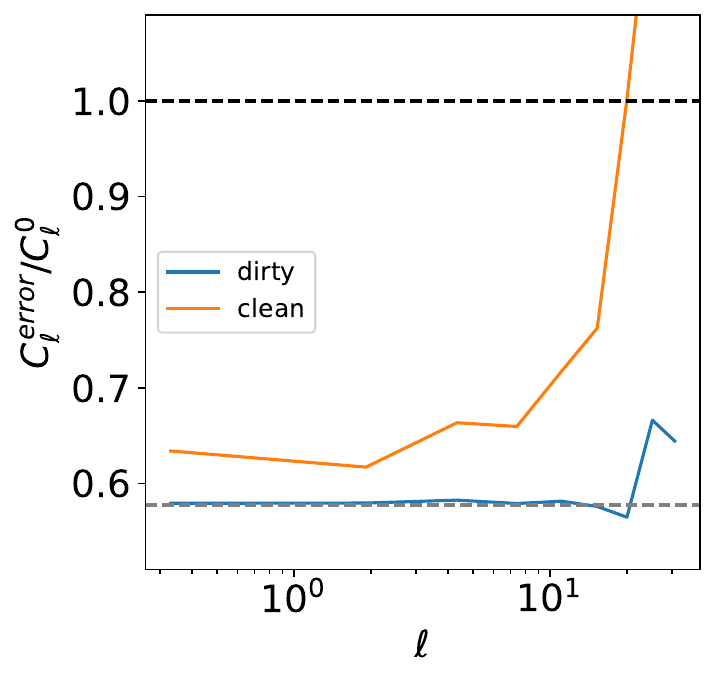}
    \caption{The ratio $C_\ell^{error}/C_\ell^{0}$ as a function of $\ell$ for random instrumental phase errors of $\Delta\phi_0=50^\circ$ on the level of dirty maps (blue) and clean maps (orange), respectively. The grey dashed line shows the theoretical estimation on the level of dirty maps. The ratio on the level of clean maps is roughly consistent with that on the level of dirty maps at very large scales but overestimated at small scales. This implies the coupling between the impacts of phase errors and the systematics from deconvolution on clean maps.}
    \label{fig:deconv}
\end{figure}

In this section, we present the coupling between the systematics from deconvolution and the impacts of phase errors on the angular power spectrum.
With the incomplete {\it uvw} coverage due to the orbit configurations and the shading of the Moon, the pseudo inverse of the dirty beam inevitably induces extra systematics when we deconvolve dirty maps to clean maps. We plot the ratio $C_\ell^{error}/C_\ell^{0}$ in Figure~\ref{fig:deconv} on the level of dirty maps and clean maps respectively at 30\,MHz. We set the regularization parameter $\epsilon$ to be $10^{-6}$. Here we add random instrumental phase errors with $\Delta \phi_0=50^\circ$ and reconstruct the sky map with NSIDE=16. We find that after deconvolution the ratio $(C_\ell^{error}/C_\ell^{0})$ on clean maps is roughly consistent with that on dirty maps at large scales but overestimated at small scales. This shows that the impacts of phase errors on clean maps are coupled with the systematics from deconvolution. It is not easy to separate them, which is not the scoop of this paper either. Therefore, in this work, we mainly perform analysis on the level of dirty maps, which can reflect the impacts of phase errors in a more reliable way. 

\begin{figure*}
    \centering
    \includegraphics[height=0.8\columnwidth]{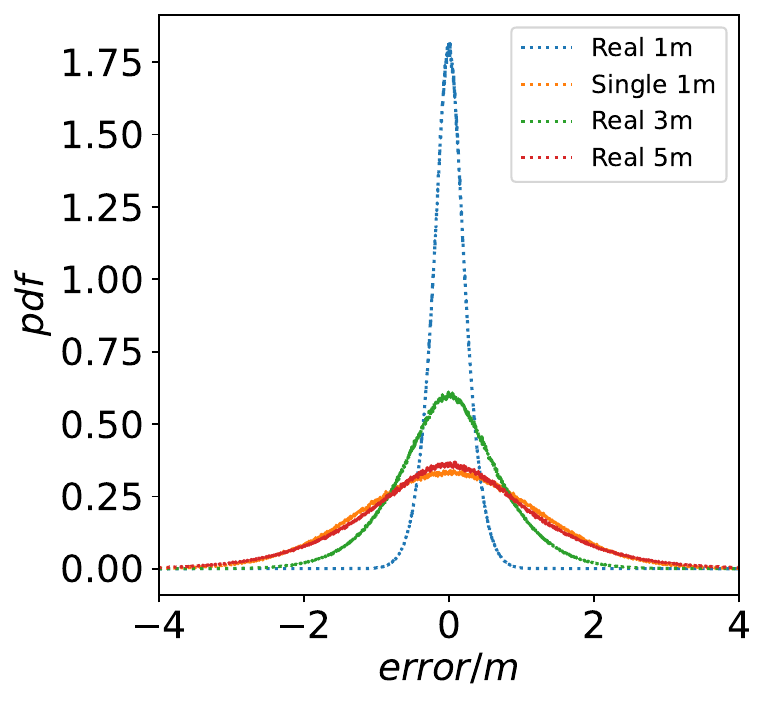}
    \includegraphics[height=0.8\columnwidth]{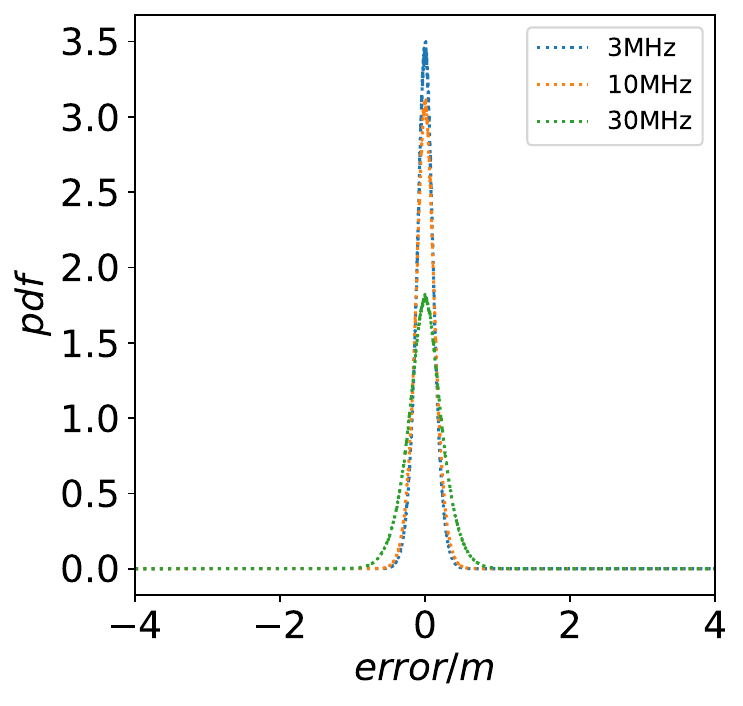}
    \caption{The distribution of effective baseline errors. We vary the baseline error model of $\Delta r_0=1$\,m (blue for ``Real'' model and orange for ``Single'' model), $3$\,m (green), and $5$\,m (red) at $f=30$\,MHz in the left panel, while vary the frequency of $f=3$\,MHz (blue), $10$\,MHz (orange), and $30$\,MHz (green) with the ``Real'' model of $\Delta r_0=1$\,m in the right panel.}
    \label{fig:bl_error}
\end{figure*}

\section{The theoretical estimation for the geometric phase errors}
\label{appendix_b}
In this section, we detail the theoretical estimation of $C^{error}_\ell/C_\ell^0$ for geometric phase errors. Assuming the effective baseline error $\Delta \mathbf{r}^{\rm eff}_{\alpha}$ is isotropic, $\hat{\mathbf{n}}\cdot\Delta \mathbf{r}^{\rm eff}_{\alpha}$ should be direction-independent and equivalent to the tangential component of the effective baseline errors along any direction. We thus use the distribution of one tangential component to estimate $C^{error}_\ell/C_\ell^0$, as plotted in Figure~\ref{fig:bl_error}. We vary the baseline error model at $f=30$\,MHz in the left panel, while vary the frequency with the fixed ``Real'' model of $\Delta r_0=1$\,m in the right panel. Note that in the case of ``Real'' model, the effective baseline errors $\Delta \mathbf{r}^{\rm eff}_\alpha$ depend on baseline lengths and frequencies. In principle, short baselines at low frequencies should have more measurement counts $N_{\rm eff}$ so that their effective errors should be smaller and their impacts should be weaker. Therefore, we divide baselines into 5 bins based on their lengths, evaluate the theoretical value of $C^{error}_\ell/C_\ell^0$ for each bin, and then average them with a weight determined by the covariance matrix of dirty maps reconstructed with data from different baseline bins. We plot the averaged results with the dashed lines in Figure~\ref{fig:all-sky geo} and find that they are consistent with the simulated results.


\bibliography{dslerr}{}
\bibliographystyle{aasjournal}



\end{document}